\documentclass[]{aa}
\usepackage{graphicx}
\usepackage{amsfonts}
\newcommand{\logp}{\ensuremath{\log\mbox{P}\ }}

\newcommand{\kms}{\ensuremath{\mbox{ km}\mbox{ s}^{-1}\ }}

\newcommand{\msun}{\mbox{ M}_\odot}

\begin{document}

\title{The formation and evolution of binary systems.\\ III. Low-mass
  binaries in the Praesepe cluster.\thanks{Based on observations obtained at
    the Canada-France-Hawaii Telescope}}

\author{J. Bouvier\inst{1}\and G. Duch\^ene\inst{1,2} \and
  J.-C. Mermilliod\inst{3} \and T. Simon\inst{4}}

\offprints{J. Bouvier}
\mail{jbouvier@laog.obs.ujf-grenoble.fr}

\institute{Laboratoire d'Astrophysique, Observatoire de Grenoble,
  Universit\'{e} Joseph Fourier, B.P.  53, 38041 Grenoble Cedex 9, France\\
  http://www-laog.obs.ujf-grenoble.fr/ \and UCLA Division of Astronomy and
  Astrophysics, Los Angeles, CA 90095-1562, USA \and
  Institut d'Astronomie de l'Universit\'e de Lausanne, CH-1290 
Chavannes-des-Bois, Switzerland \and Institute for Astronomy, University of Hawaii, 2680
  Woodlawn Drive, Honolulu, HI 96822, USA}

\date{Received / Accepted }

\authorrunning{J. Bouvier et al.}
\titlerunning{Praesepe low-mass binaries}

\abstract { With the aim of investigating the binary population of the
  700~Myr old Praesepe cluster, we have observed 149 G and K-type cluster
  members using adaptive optics. We detected 26 binary systems with an
  angular separation ranging from less than 0.08 to 3.3 arcsec (15--600
  AU).  After correcting for detection biases, we derive a binary frequency
  (BF) in the $\logp$ (days) range from 4.4 to 6.9 of 25.3$\pm$5.4\%, which
  is similar to that of field G-type dwarfs (23.8\%, Duquennoy \& Mayor
  1991).  This result, complemented by similar ones obtained for the 2~Myr
  old star forming cluster IC 348 (Paper II) and the 120~Myr old Pleiades
  open cluster (Paper I), indicates that the fraction of long-period
  binaries does not
  significantly evolve over the lifetime of galactic open clusters.\\
  We compare the distribution of cluster binaries to the binary populations
  of star forming regions, most notably Orion and Taurus, to critically
  review current ideas regarding the binary formation process. We conclude
  that it is still unclear whether the lower binary fraction observed in
  young clusters compared to T associations is purely the result of the
  early dynamical disruption of primordial binaries in dense clusters or
  whether it reflects intrinsically different modes of star formation in
  clusters and associations. We also note that if Taurus binaries result
  from the dynamical decay of small-N protostellar aggregates, one would
  predict the existence of a yet to be found {\it dispersed\/} population
  of mostly single substellar objects in the Taurus cloud.
   }

\maketitle

\section{Introduction}

Binary and multiple systems provide a fossil record of the star formation
process. In the last decade, studies of various galactic populations (field
stars, open clusters, star forming regions) have led to the conclusion that
most solar-type stars occur in binary systems rather than in isolation
(e.g., Duquennoy \& Mayor 1991, Mathieu et al. 2000). Hence, the most
common output of protostellar collapse appears to be the formation of
multiple systems. Beyond this indisputable observational result, the way
stellar systems form remains an important issue (Bodenheimer et al. 2000),
and so a robust determination of the detailed properties of binary stars, 
e.g., the distributions in their orbital periods, mass-ratios, and orbital 
eccentricities, is critically
needed to guide theoretical models (Clarke et al. 2001, Ghez 2001).

Equally important in order to get clues to the star formation process is to
determine whether the properties of binaries depend upon the environment
in which they form, and whether these properties evolve over time or, to the
contrary, remain stable during pre-main sequence and main sequence
evolution: that is, are the statistical properties of binary populations 
universal and do they unambiguously reflect the processes which gave them 
birth, or do they vary both over time and from place to place in the solar
neighbourhood? In order to address these issues, multiple systems have to
be sampled and characterized in various types of environments and in
stellar populations that have reached different stages of evolution.

A number of studies have been devoted to these issues. Large scale searches
for binaries have been completed among low-mass field dwarfs (Duquennoy \&
Mayor 1991, Fisher \& Marcy 1992, Tokovinin 1992), T Tauri stars in young
stellar associations (e.g., Leinert et al. 1993, Ghez et al. 1993, Simon et
al. 1995) and low-mass stars in young clusters (e.g., Bouvier et al.  1997,
Duch\^ene et al. 1998, Patience et al. 1998, Mermilliod \& Mayor 1999). The
results indicate that in most surveyed regions the low-mass stars exhibit a
similar fraction of close visual binaries with semi-major axes between a
few tens to about 1000 AU (see Duch\^ene 1999 for a summary). There are,
however, two notable exceptions.  First, the Taurus star-forming
association appears to contain about twice as many binaries as are present
in the field (Leinert et al. 1993, Ghez et al.  1993).  Second, there is a
marked deficit of wide binaries, with semi-major axes in the range from
1000 to 5000 AU, in the Orion cluster as compared to the field binary
population (Scally et al. 1999).

One of the difficulties in interpreting these results is that the various
studies were performed with different techniques, thus introducing
different observing biases (see Duch\^ene 1999). Motivated by the need to
obtain homogeneous data sets for close visual binaries, especially in
cluster environments whose study has been somewhat neglected compared to
star forming regions and field populations, we have started high angular
resolution surveys of low-mass stars in nearby young clusters. For this
project, we have selected a number of clusters having ages in the range
from 2~Myr to 700~Myr, in order to investigate any evolution of the binary
content with time. The first 2 papers of this series reported on the
results obtained for the $\sim$100 Myr Pleiades cluster (Bouvier et al.
1997, Paper I), the $\sim$2 Myr IC 348 cluster (Duch\^ene et al. 1999,
Paper II), while preliminary results for the $\sim$80~Myr Alpha Per
cluster have been reported by Eisl\"offel et al.  (2001).

We report here the results obtained on the binary population of the 700~Myr old
Praesepe cluster from high angular resolution observations of 149 low mass
cluster members. Section 2 describes our data aquisition and analysis techniques, 
which are similar to those used in Papers I and II. The binary frequency among
the low-mass stars in Praesepe and the binary properties of the cluster are 
derived in Section 3.
Combining these data with those available in the literature for other young
clusters, star forming associations, and the field, we discuss in Section 4
the implications of the trends (or lack of) observed in binary frequency as
a function of time and environmental conditions for the binary formation
process. 

\section{Observations and data analysis}

The sample was primarily drawn from the proper motion study of Praesepe
candidate members by Jones \& Stauffer (1991, hereafter JS91). From this
catalogue, we first selected 70 candidates in a B-V range between 0.52 and
1.4 which had a proper motion membership probability larger than 90\%.
These candidates were observed in February and December, 1997. For another
observing session in January 1998, 53 additional candidates were selected
from JS91 in the same B-V range, which had a proper motion membership
probability larger than 75\% and visual photometry consistent with
membership in a (V,B-V) color-magnitude diagram. We completed our sample
with 26 additional candidates having both radial velocities and photometry
consistent with Praesepe membership, which were selected from Mermilliod's
Open Clusters Database (WEBDA, Mermilliod 1999) and amongst Praesepe halo
candidates (Mermilliod et al., 1990).

%
%
%
%
%

A total of 149 Praesepe G and K dwarfs were thus observed in 1997 and 1998
at the Canada-France-Hawaii Telescope with the adaptive optics system PUEO
(Rigaut et al. 1998). The IR camera was MONICA (Nadeau et al. 1994) in Feb.
1997 and KIR (Doyon et al.  1998) in Dec. 1997 and Jan. 1998, which provide
a field of view of 9{\arcsec} and 36{\arcsec}, respectively. We used the
same aquisition procedure for all runs. The targets were observed
successively in four quadrants of the camera, for a total integration time
of typically 60s; the 4 exposures were subsequently registered and added to
produce a final image that was properly sky subtracted and flat-field
corrected. The primaries were first observed in either the H or K band,
depending on the atmospheric turbulence conditions, in order to optimize
the adaptive correction. Whenever a binary was detected in real-time, it
was observed in at least one other filter in order to subsequently check
the membership of the companion in a color-magnitude diagram. During each
run, the images produced by the adaptive optics system were
diffraction-limited in the H and K bands, providing a spatial resolution of
0.09{\arcsec} and 0.13{\arcsec} FWHM, respectively.

Aperture photometry was performed using IRAF/APPHOT and calibrated using
several UKIRT Faint Standards observed during the 3 runs. The photometric
accuracy is of order of 0.05 magnitudes in the JHK bands.  Table~1 lists
the near infrared magnitudes of non resolved primaries while those of
Praesepe binaries are listed in Table~\ref{table_bin}. For the binaries,
differential photometry was obtained by fitting a template PSF
simultaneously to the primary and the companion within IRAF/DAOPHOT.  The
PSF was provided by unresolved Praesepe stars that were observed just
before and/or just after the exposure on the binary. Several PSF templates
were used for each binary, thus providing an estimate of the photometric
error. Typically, the differential photometry is accurate to within 0.02
magnitudes, but the error may be larger for binaries with a separation
close to the resolution limit or for companions close to the detection
limit. The magnitude difference between the companion and the primary was
combined with the aperture photometry of the system to provide the
magnitude of each component.

The separation and position angle were derived from the photocenter of the
components in the image as provided by the PSF fitting algorithm. The plate
scale and orientation of the detector were calibrated by observing IDS
astrometric standards (Van Dessel \& Sinachopoulos, 1993) during each run.
The rms error is typically 5 mas on the separation and 0.1{\degr} on the
position angle.

\begin{table*}
  \caption{Photometry of non resolved Praesepe primaries$^1$}
  \label{table_single}
  \smallskip
\begin{tabular}{llll|llll|llll|llll}
\hline
BDA    &Filt  & Mag  &Run &BDA    &Filt  & Mag  &Run &BDA    &Filt  & Mag  &Run &BDA    &Filt  & Mag  &Run \\
\hline
9 & K & 9.51 & jan98 & 23 & K & 9.64 & feb97 & 30 & K & 9.75 & jan98 & 48 & K & 10.17 & jan98 \\ 
49 & H & 9.27 & jan98 & 58 & K & 9.64 & feb97 & 70 & H & 10.08 & dec97 & 127 & H & 9.41 & jan98 \\ 
141 & H & 10.27 & dec97 & 162 & H & 9.15 & feb97 & 172 & K & 10.34 &
jan98 & 181 & H & 9.04 & jan98 \\ 
 ... & ... & ... & ...& ... & ... & ... & ...& ... & ... & ... & ...& ... & ... & ... & ... \\
\hline
\multicolumn{16}{l}{$^1$ Table~1 is available electronically in extenso at CDS, Starsbourg}\\
\end{tabular}
\end{table*}

\begin{table*}
  \caption{Astrometric and photometric properties of Praesepe binaries}
  \label{table_bin}
  \smallskip
\begin{tabular}{lrrrclllllllllcl}
\hline
BDA    &KW$^a$     &VL$^b$      &JS$^c$     &V     &B-V  &sep    &P.A.
&J$_{AB}$      &H$_{AB}$      &K$_{AB}$   &$\Delta$J  &$\Delta$H   &$\Delta$K   &q &Notes\\
&&&&&& \arcsec & \degr \\
\hline
79 &79 &572  &211 &12.10 &0.91 &0.174 &73.5 &10.40 &9.98 &9.83 & &1.40 &1.58 &0.64 &1\\
90 &90 &598  &222 &10.89 &0.70 &0.184 &27.4 &     &9.20 &9.14 & &0.90 &1.02 &0.75 &1\\
100 &100 &621  &228 &10.55 &0.58 &1.03 &97.3 &9.53 &9.22 &9.25 &     &6.20 &5.70 &0.11\\
164 &164 &789  &279 &11.31 &0.70 &0.256 &18.4 &     &     &9.60 &     &     &2.76 &0.42\\
198 &198 &870  &306 &12.62 &0.97 &1.153 &354.48 &10.82 &10.37 &10.25 &2.64 &2.50 &2.35 &0.47\\
275 &275 &993      &     &9.96 &0.58 &0.21 &90.4$^\dagger$ &8.83 &8.60 &8.54 &0.03 &-0.02 &0.02 &1.0 &1\\
287 &287 &1014      &     &10.37 &0.59 &1.798 &24.84 &9.46 &9.07 &9.15 &5.63 &5.31 &5.13 &0.15 &2\\
297 &297 &1033  &362 &11.64 &0.86 &0.126 &188.3 &     &9.49 &9.40 & &0.61 &0.56 &0.83 &3\\
322 &322 &1070  &375 &10.87 &0.68 &$<$0.09 &$\sim$150 &     &9.20 &9.14
&     &     &$\sim$0.5 &0.8 &4\\
334 &334 &1091  &387 &11.01 &0.72 &0.091 &44.9 &9.56 &9.20 &
&1.14 &0.85 &     &0.77 &1\\
365 &365 &1142      &407 &10.18 &0.65 &0.373 &109.27 &8.86 &8.52 &
&0.87 &0.83 &     &0.79 &5\\
401AB &401 &1214      &436 &12.97 &1.00 &1.688 &239.56 &10.77 &10.23 &10.15 &2.51 &2.41 &2.30 &0.43\\
401AC$^*$ &401 &1214      &436 &12.97 &1.00 &1.776 &286.69 &10.77 &10.23 &10.15 &     &6.1 &5.4 &0.10\\
466$^{**}$ &466 &1345  &486 &10.99 &0.65 &2.184 &304.97 &9.71 &     &     &4.48 &4.12 &3.87 &0.26\\
488 &488 &1399  &509 &11.43 &0.73 &1.263 &198.1 &     &9.76 &9.73 &     &4.91 &4.65 &0.17\\
495 &495 &1416  &515 &9.97 &0.66 &0.072 &159.3$^\dagger$ &9.66 &9.35
&     &     &0.03 &&0.99 &6\\
533  &533 &237  &122 &11.59 &0.90 &0.124 &16.2 &     &9.54 &     &
&0.25 &     &0.93 &7\\
540 &540 &387 &167 &11.03 &0.69 &3.35 &233.59 &9.69 &9.33 &9.29 &6.12
&5.88 &5.67 &0.11 &8\\
809 &     &          &194 &13.04 &1.13 &1.737 &325.96 &10.99 &10.44 &10.32 &1.62 &1.49 &1.35 &0.66\\
901 &     &          &354 &13.82 &1.39$^\S$ &2.428 &191.7 &     &10.66 &10.52 &     &0.34 &0.31 &0.92\\
1184 &     &184  &102 &11.85 &0.79 &1.285 &151.95 &     &9.85 &9.73 &
&2.50 &2.43 &0.48 &9\\
1452 &     &452      &186 &13.87 &1.32$^\S$ &0.390 &130.9 &     &10.80 &10.68 &     &2.78 &2.62 &0.38\\
1995 &     &995      &350 &12.97 &1.21 &0.337 &289.49 &10.71 &10.11 &10.01 &0.18 &0.11 &0.18 &0.73\\
2029 &     &1029      &359 &12.90 &1.04 &0.520 &204.9 &     &10.46 &10.38 &     &3.94 &3.78 &0.23\\
2085 &     &1085      &383 &12.89 &1.10$^\S$ &0.642 &186.9 &     &10.54 &10.44 &     &4.00 &3.79 &0.22\\
2418 &     &1418      &516 &13.67 &1.18 &0.394 &221.5 &     &10.84 &10.65 &     &3.1 &3.16 &0.32\\
2692 &     &1692  &588 &10.92 &0.72 &0.405 &234.6 &     &9.17 &9.13 &     &0.50 &0.61 &0.85\\
3231 &     &          &231 &13.31 &1.39$^\S$ &0.168 &172.0$^\dagger$ &     &10.26 &10.16 &     &0.13 &0.09&0.97 \\
\hline
\multicolumn{16}{l}{$^a$ Klein-Wassink 1927, $^b$ Vanderlinden 1933, $^c$
  Jones \& Stauffer 1991}\\
\multicolumn{16}{l}{$^\S$ Photographic magnitudes.}\\
\multicolumn{16}{l}{$\dagger$ Within 3$\sigma$ photometric errors, the
  P.A. could be 180\degr\ away from the listed value.}\\
\multicolumn{16}{l}{$^*$ BDA 401C is a probably a field object.}\\
\multicolumn{16}{l}{$^{**}$ Due to non-photometric conditions, only
  differential photometry was obtained in the H and K bands.}
\end{tabular}
Notes: $^1$~Photometric binary; $^2$~SB1, P=7635d; $^3$~SB1, long period; 
$^4$~SB1, P$>$10000d;  
$^5$~Triple, A: SB1O, B:single; $^{6}$~Triple, A:
single, B: SB2; $^7$~SB2, long period; $^8$~SB1, P=1149d; $^9$~SB1, P=1.23d. 
\end{table*}

\section{Results}

We detected 26 binaries and one triple system having separations less than
7{\arcsec} among 149 Praesepe G and K dwarfs. Table~\ref{table_bin}
provides their names and cross-identifications, visual photometry, and
indicates whether they were previously known as either photometric or
spectroscopic binaries. 

\begin{figure*}[th]
  \begin{center}
    \leavevmode
    \includegraphics[width=0.9\hsize]{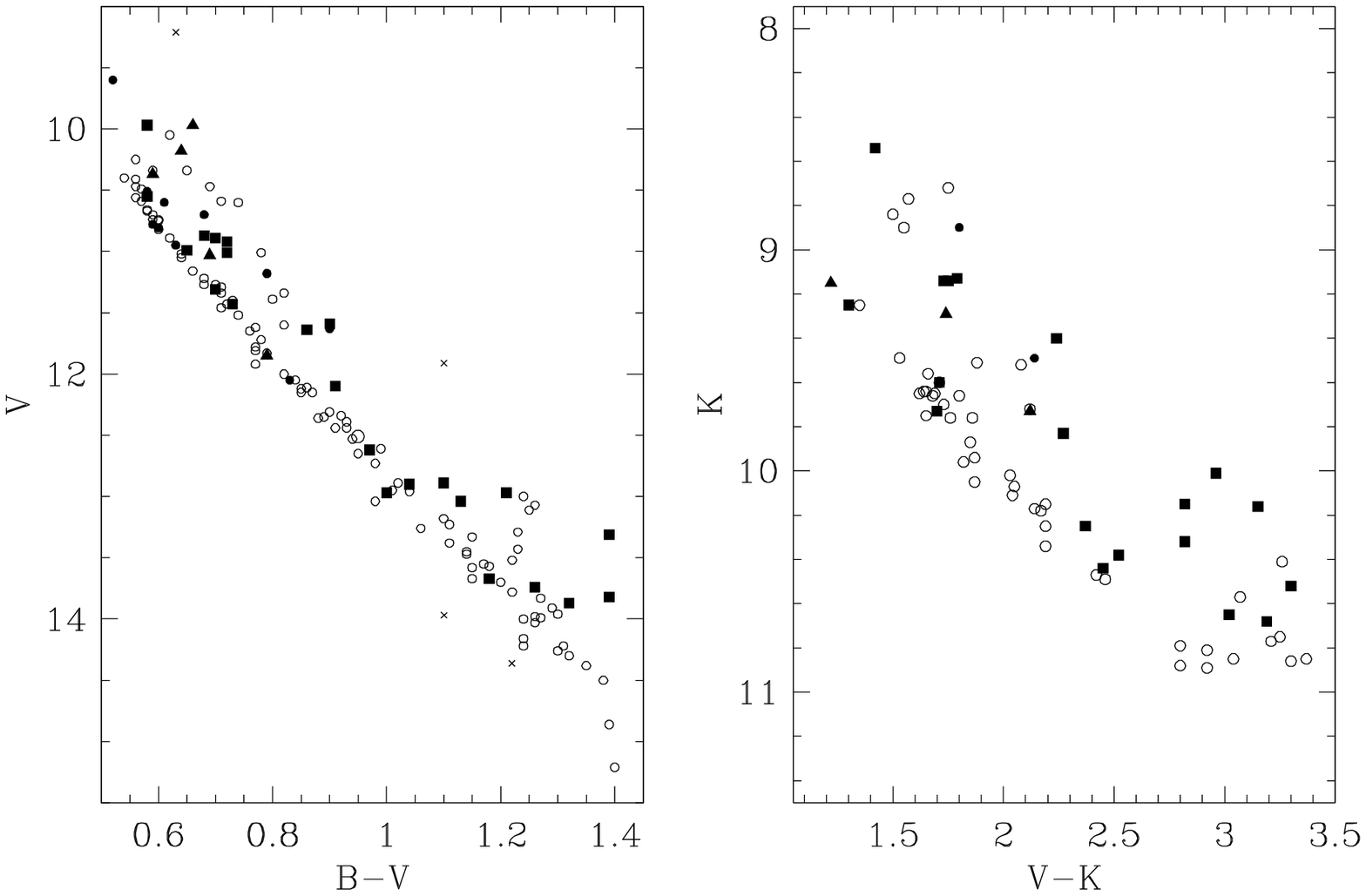}
  \end{center}
  \caption{Color-magnitude diagrams (CMD) for low-mass Praesepe members.  
    The (V, B-V) CMD ({\it left panel}) contains the whole sample observed
    with PUEO, while the (K, V-K) one ({\it right panel}) contains only a
    subset of it since some primaries were not observed in the K-band.
    Symbols are as follows: {\it filled squares}: binary systems resolved
    with PUEO, {\it filled dots}: known spectroscopic binaries, {\it filled
      triangles}: triple systems resolved by PUEO, {\it open circles}:
    ``single stars'', {\it crosses\/}: suspected non members. Several
    systems resolved in the near-infrared appear to be single in the
    visible.}
  \label{cmd}
\end{figure*}

\subsection{Comparison with previous work} 

Of the 26 binaries reported here, eight were previously known as
spectroscopic binary (SB) systems (Mermilliod \& Mayor 1999): three
long-period SBs (BDA 297, 322, 533) are resolved here with a projected
separation of 23 AU or less, while three others have too short a period to
be resolved (BDA 287, 540, 1184) and so the companion reported here makes them
triple systems. The two remaining systems, BDA 365 and BDA 495, are known
to be spectroscopic triples (Mermilliod et al.  1994). They both consist of
a long-period binary with one of the components itself being a short-period
binary. Both objects are spatially resolved here, and, given their angular
separations and spectroscopic orbits, it is likely that we resolved the
widest binary pair of these systems.


Figure~\ref{cmd} shows a (V, B-V) color-magnitude diagram (CMD) for the
whole sample of G and K Praesepe members observed in this survey.  The
different symbols indicate previously known spectroscopic binaries, visual
binaries detected here by adaptive optics, and ``single'' stars.  Not
surprisingly, the sample contains a number of (presumably short-period) SBs
located on the main and binary sequences of the cluster that are not
resolved by adaptive optics. More interestingly, about 15 objects that are
not detected as binaries, either from spectroscopy or adaptive optics, lie
more than 0.5 mag above the main sequence of the cluster and must therefore
be nearly equal-mass binary systems. In order to have escaped detection,
these systems must have an orbital period in the range from about 3 10$^3$
to 3 10$^4$ days, i.e., a semi-major axis in the range from about 4 to 15
AU.  Finally, although many of the binaries that are resolved in the
present study are displaced above the main sequence as expected, about a
third of them are located on the cluster main sequence and were not
previously detected as binaries through photometry. BDA 1184 (the triangle
at V = 11.85, B-V = 0.79) is the most extreme case: it appears to be a
single star but is in fact a triple system, a short-period spectroscopic
binary and a wider companion at 1$\farcs$28. This simply means that many
faint and red companions are hidden in visible light and are too
light-weight to show up in radial-velocity observations.

A (K, V-K) CMD, also shown in Figure~\ref{cmd}, is more appropriate to
detect very red companions photometrically. In this diagram, BDA 1184 shows
a vertical displacement of 0.42 mag. Another example, BDA 2418, which lies
right on the single star sequence in the (V, B-V) plane, already shows a
displacement of 0.15 mag in (I, V-I) and 0.35 mag in (K, V-K). However, the
star BDA 287, another triple system, is still located on the single star
sequence even in (K, V-K), because the K magnitude difference is 5.13. 

To conclude this section, the common use of (V, B-V) colour-magnitude
diagram to study binarity in open clusters from photometric data is
not a good choice or strategy. (I, V-I) planes are certainly better and JHK
observations provide still much more information on faint red companions.
Observations through the $BVIK$ filters would therefore permit the
detection of a wealth of new binary candidates, which would raise the
binary frequency to values that are probably more realistic. However, 
direct observations are still needed to detect systems with large magnitude
differences or with a multiplicity of order higher than 2.


\subsection{Binary frequency and orbital period distribution} 

Table~\ref{table_bin} also lists the astrometric and infrared properties of
the systems and their components. Although BDA 322 is clearly elongated on
the images, its separation is too small for deriving precise 
astrometric and photometric properties, and so we list only approximate 
values in Table~\ref{table_bin}. The masses of the individual binary components
and the resulting mass ratios, $q=M_2/M_1$, were derived from the JHK
photometry of the components using the mass-magnitude relationships from
Baraffe et al.  (1998) models, assuming an age of 0.7 Gyr and a distance
modulus for Praesepe of $(m-M) = 6.28$ ($d=180$pc, Robichon et al. 1999).
For the primaries or secondaries that are themselves unresolved spectroscopic
binaries, this method would overestimate their mass by up to about 20\% for
equal-brightness components.

\begin{table*}
\caption[]{Binary frequency (see text)} 
\label{bf}
\begin{flushleft}
\begin{tabular}{lllllll}
\hline
$\log P_{orb}$                 &4.4-4.9    &4.9-5.4 &5.4-5.9 &5.9-6.4  &6.4-6.9&days\\
Semi-major axis                &18-39      &39-86   &86-183  &183-394  &394-852&AU\\
Sep. range                      &0.08-0.17  &0.17-0.38 &0.38-0.81  &0.81-1.74&1.74-3.76 &{\arcsec}\\
$\Delta m_{max}$                &2.2       &3.0     &4.5      &6.7      &7.2&mag\\ 
q$_{min}$                       &0.50      &0.40    &0.20     &$<$0.1   &$<$0.1\\ 
N$_{detect.}$ [$q_{min},1.0$]   &4         &6       &5        &6        & 4        \\ 
N$_{undetect.}$[$0.1,q_{min}$]  &6.4      &5.4     &0.95      &0        & 0    \\ 
N$_{tot}$                       &10.4      &11.4     &5.95      &6        & 4   \\
\hline
{\bf B.F. Praesepe} (rms)      &7.0 (3.5)  &7.7 (3.1)&4.0 (1.8)&4.0 (1.6)&2.6 (1.3) &\%\\
{\bf B.F field G dwarfs}       &5.4        &5.3      &5.0      &4.5    & 3.9  &\%\\
\hline
$\log P_{orb}$=4.4-6.9 & \multicolumn{3}{l}{{\bf B.F. Praesepe:} 25.3 $\pm$
  5.4\%} & \multicolumn{3}{l}{{\bf B.F. field:} 23.8\%}\\
\hline
\end{tabular}
\end{flushleft}
\end{table*}

In order to ascertain photometric membership, we plotted the primaries and
secondaries in various JHK color-magnitude diagrams, comparing their
location in these diagrams with the 700 Myr isochrone from Baraffe et al.
(1998). The primaries and secondaries are all consistent with being
Praesepe members within the photometric errors. However, the third
component of the BDA 401 system, BDA 401C, lies far away from the isochrone
and is probably a field object. We therefore ignore it in the following
discussion and consider BDA 401 to be a double.  Finally, a few binaries were
observed at only one wavelength, but these systems are so tight that their 
binary nature is not in serious doubt.



\begin{figure}[th]
  \begin{center}
    \leavevmode
    \includegraphics[width=0.9\hsize]{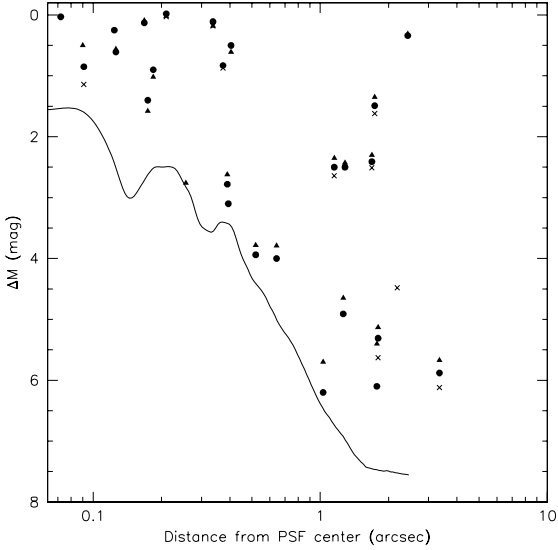}
  \end{center}
  \caption{Limit of detection for faint companions. The location of resolved
    Praesepe binaries in this diagram is indicated (crosses: $\Delta J$,
    filled dots: $\Delta H$, filled triangles: $\Delta K$). The curve
    indicates the maximum magnitude difference detectable on AO images at
    any distance from the center of the primary. It was derived by
    computing the 5$\sigma$ noise level on radial profiles of unresolved
    Praesepe primaries. By adding artificial companions to the primaries on
    the images, we empirically verified that this curve corresponds to the
    limit of detectability of faint and/or close companions. }
  \label{detect}
\end{figure}

That we find only one chance projection at a distance of less than
7{\arcsec} in a sample of 149 stars is consistent with the 2MASS Point
Source Catalogue Statistics, which predict about 1200 objects per square
degree down to a magnitude of K=15 in the direction of Praesepe. This
translates into $\sim$0.015 objects within a 7{\arcsec} radius and leads 
to an estimate of 2 chance projections in the present sample.




\begin{figure}[th]
  \begin{center}
    \leavevmode
    \includegraphics[width=0.9\hsize]{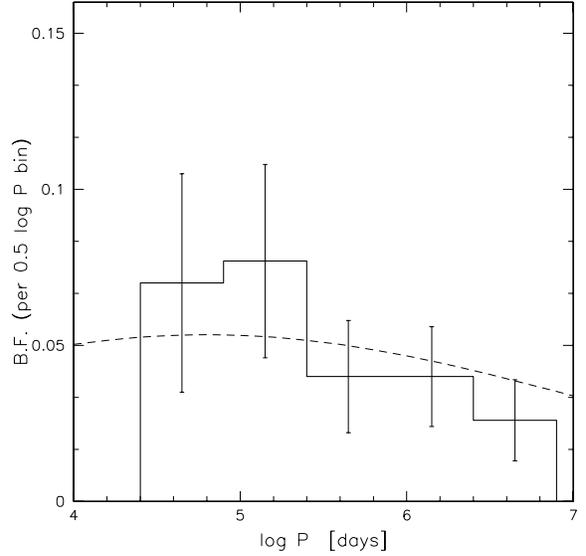}
  \end{center}
  \caption{Distribution of orbital periods for Praesepe binaries
    (histogram). The error bars represent Poisson noise (see text). The
    dashed curve is the orbital period distribution of field G dwarf
    binaries as derived by DM91.}
  \label{porb}
\end{figure}

In order to estimate the binary frequency among the Praesepe G and K dwarfs, 
a correction factor has to be applied to the number of systems actually
detected to account for the detection limit of our survey. The largest
magnitude difference we are able to detect between the secondary and the
primary is shown in Figure~\ref{detect} as a function of angular
separation. At separations close to the diffraction limit, the detection of
faint companions is limited by the constrast against the bright primary
while, at large separations, the detection is background limited. We
therefore proceed to derive a correction factor in various separation bins
as follows.

In each bin, we convert the maximum reachable contrast between the
secondary and the primary, $\Delta m_{max}$, into a minimum mass ratio
$q_{min}$ using the mass-magnitude relationship from the Baraffe et al.
(1998) models. We assume that the mass-ratio distribution of the Praesepe
binaries is the same as that of the field G dwarfs derived by Duquennoy \&
Mayor (1991, DM91). This assumption is consistent with the mass ratio
distribution derived for our binary sample. While the overall
q--distribution is rather flat (see Table~\ref{table_bin}), restricting the
analysis to the separation range where we can detect all companions down to
$q=0.1$ (sep$\geq$0.8\arcsec, see Table~\ref{bf}) so that we can compare to
DM91, we count 10 companions. Mass ratios in this small subsample range
from 0.11 to 0.92, with a mean of 0.38. In DM91 survey, for binaries with
periods longer than 10$^4$~days, the average mass ratio is 0.40. Both
results are very similar, though ours admitedly relies on only a few
systems. The fraction of missed companions is then found by integrating
DM91's mass-ratio distribution between q=0.1 and q=$q_{min}$. We finally
apply this correction factor to the number of detected binaries to obtain
the total number of systems with mass-ratios larger than 0.1 in this
separation range. The 1$\sigma$ uncertainties on BF in each $\logp$ bin
correspond to Poisson noise, i.e., $\sigma = \sqrt{N_d} \times (1 +
N_u/N_d) $, where $N_d$ and $N_u$ are the number of detected and missed
systems, respectively (see Table~\ref{bf}).

In order to establish the distribution of orbital periods, i.e., the
frequency of binary systems in each $\logp$ bin, we convert the intervals
of projected separation into bins of orbital periods. Angular separations
$\rho$ are statistically corrected for projection effects to yield the
semi-major axis according to: $\log a = \log (\rho\times d) + 0.1$ (DM91), 
where $d=180$~pc is the distance to the Praesepe cluster. Kepler's 3rd law 
with an average mass of 1.3 $M_\odot$ for the system yields the corresponding 
range of orbital periods (see Table~\ref{bf}).

The last lines of Table~\ref{bf} list and also Figure~\ref{bf} illustrates
the derived frequency of Praesepe low-mass binaries in each $\logp$ bin
between 4.4 and 6.9 (P in days), as compared to the frequency of G dwarf
binaries over the same separation range\footnote{As in our previous paper
  on the Pleiades (Bouvier et al. 1997), we define the binary frequency,
  B.F., as the number of binary orbits divided by the number of primaries
  in the sample. This is equivalent to the companion star fraction, $csf =
  (B+2T)/(S+B+T)$, where S, B, and T are single, binary, and triple
  systems, respectively.}.  Even though the statistical
uncertainties in each bin are somewhat large, especially at the shortest
orbital periods where the incompleteness correction is significant, the
overall binary frequency appears to be very similar among the solar-type
Praesepe stars and the field G dwarfs, amounting to 25.3$\pm$5.5\% and
23.8\%, respectively, for the 4.4-6.9 range in $\logp$.
 Restricting the comparison to the range of orbital
periods not affected by detection biases ($\logp$ from 5.4 to 6.9), the BF
amounts to 10.7$\pm$2.6\% for Praesepe stars compared to 13.4\% for field
dwarfs. We thus conclude that the frequency of long period binaries among
Praesepe G and K stars is indistinguishable from that measured among the G
field dwarfs by DM91. This conclusion seems to apply
to short-period systems as well ($\logp\leq 4.0$) as Mermilliod's \& Mayor's
(1999) investigation of spectroscopic binaries in the Praesepe cluster
yields BF$ = 25\pm 5\%$, as compared to 21\% for field dwarfs.


Praesepe has nearly the same age as the Hyades cluster. Unfortunately,
comparison of the BF between the two clusters is limited by the fact that
there is only a slight overlap in the separation ranges covered by Patience
et al.'s (1998) survey in the Hyades and ours in Praesepe. In the 15--50 AU
common range of semi-major axes, we detect 8 companions, i.e. an observed
companion fraction of 5.3$\pm$1.9\%. In the same separation range, Patience
et al. (1998) detected 9 companions, out of which 3 would not have been
detected in our survey given their flux ratios. This amounts to a companion
star fraction of 3.7$\pm$1.5\%, less than 1$\sigma$ lower than our Praesepe
estimate.  Within this restricted range of orbital periods, we thus do not
find any significant difference between Hyades and Praesepe binary
fraction, though this conclusion is obviously based on small number
statistics.

\section{Discussion}

In this section, we first briefly review current observational results on
the statistics of binary frequency in young clusters and associations,
including results from both this and previous papers of this series and
other published work. The comparative analysis of the properties of binary
populations in different environments and at various evolutionary stages
provides constraints on binary formation models, which are discussed below
with the emphasis on the difficulties that are encountered by the various
formation scenarios. The discussion is restricted to visual (a $\sim$
20-1000 AU), low-mass binaries ($\sim$ 0.5-1.1$\msun$) that are resolved by
high angular resolution techniques.

\subsection{Binary statistics in clusters and associations}

The three young clusters we surveyed for low-mass binaries, IC 348 ($\sim$2
Myr, Duch\^ene et al. 1999), the Pleiades (120 Myr, Bouvier et al.  1997),
and Praesepe (700 Myr), have been observed with the same instrumentation,
thus resulting in similar detection biases.  Moreover, the results have
been analyzed in a consistent way, in particular with regard to the
incompleteness corrections. The three clusters are found to exhibit
identical binary frequencies over the \logp range $\sim$4.5-7.0, $<$BF$>$
$\sim$ 0.25 $\pm$ 0.05, which is also consistent with the BF measured for
solar-type field dwarfs in the same range of orbital periods (BF $\sim$
0.24, DM91). Similar results have been obtained for other clusters, Orion
($\sim$2 Myr, Prosser et al. 1994, Petr et al.  1998, Simon et al. 1999),
Alpha Persei ($\sim$ 80 Myr, Eisl\"offel et al. 1999), and the Hyades
($\sim$600 Myr, Patience et al. 1998) --- all of which exhibit a low-mass
BF in a restricted \logp range that is consistent with the field dwarf BF
when the results are analyzed in a uniform way (see Duch\^ene 1999).

Since the ages of these clusters cover a huge range from 2 to 700 Myr, the
results suggest that low-mass binaries are formed at a very early stage of
cluster evolution (at an age $\leq$ 1 Myr) and that the binary fraction
does not evolve much thereafter until the cluster eventually dissolves into
the field ($\simeq$ 1 Gyr). The lack of a significant evolution of the
binary population during the secular dynamical evolution of a cluster is
consistent with recent numerical simulations (e.g. Kroupa 2000).

Note, however, that this conclusion may not hold for more massive and/or
spectroscopic binaries. Abt \& Willmarth (1999) have reported marginal evidence
for a BF that increases with a cluster's age, from the Orion Nebula Cluster to
Praesepe, for spectroscopic binaries with A-type primaries. They interpret
this trend as the possible signature of binary formation by capture in
evolving clusters and/or preferential escape of single stars during the
secular dynamical evolution of clusters (de la Fuente Marcos 1997). No such
trend is seen for low-mass wide binaries. 

Another general result is that pairs of nearly coeval clusters (e.g., IC348
and Orion, Alpha Persei and the Pleiades, Praesepe and the Hyades) not only
exhibit similar fractions of low-mass visual binaries over the separation
range probed by adaptive optics and speckle techniques ($\sim$20-1000 AU),
but the distribution of orbital periods is consistent as well, with
admittedly large uncertainties in the shape of the $\logp$ distribution
(see Fig.~\ref{bf}). These similarities suggest that either all these
clusters were formed under very similar conditions and have evolved in the
same way, or else the binary content of clusters and their properties
depend only weakly on initial conditions.

In marked contrast with the results obtained for cluster binaries, the
binary frequency in low-density star forming associations, most notably the
Taurus-Auriga cloud, is higher by a factor of about 2 than that observed in
both clusters and field solar-type stars (Leinert et al. 1993, Ghez et al.
1993). Ghez (2001) also reported a significant difference between the \logp
distributions of binaries in clusters and those in associations, the latter
harbouring a larger fraction of wider binaries than the former.

The different properties of binary populations in clusters and associations
can be a signature of different formation mechanisms in these environments
(e.g., Durisen \& Sterzik 1995).  Alternatively, if one assumes a universal
formation mechanism that yields the same initial BF in clusters and
associations, the observed differences could reflect dynamical processes
acting very early-on, such as the rapid disruption of wide primordial
binaries in clusters through gravitational encounters (e.g. Kroupa et al.
1995). We discuss these two possibilities in turn below.

\subsection{A universal mechanism for binary formation?}

Among the various possible ways of forming low-mass binaries, tidal capture
has been shown to be inefficient even in the densest protostellar clusters
(Clarke \& Pringle 1991, Kroupa 1995, Clarke 2001) and fission of massive
protostars or protostellar disks, which could conceivably yield the
tightest binaries, seems to be prevented by the development of bar-like
instabilities (Durisen et al. 1986, Bate 1998). Therefore, multiple
fragmentation during protostellar collapse appears today to be the most
promising mechanism for creating wide multiple systems (Bodenheimer 2001).

Recent collapse calculations indicate that the likely output of multiple
fragmentation is the formation of small-N protostellar aggregates, where
N$\simeq$ 3-10 (e.g., Burkert, Bate \& Bodenheimer 1997, Klessen \& Burkert
2000). These aggregates experience rapid dynamical decay and eventually
leave a bound binary system, while other fragments are dynamically ejected
mostly as single remnants (e.g., McDonald \& Clarke 1993, 1995; Sterzik \&
Durisen 1998).  Since few-body interactions occur on a small scale within
protostellar aggregates (r $\sim$ a few 100 AU) and on a very short time
scale ($\sim 10^4$ yr), the resulting primordial binary fraction is not
expected to depend strongly upon the global properties of the star forming
region.

The fraction of primordial binaries that results from the dynamical decay
of protostellar aggregates is usually identified with the high BF observed
in loose associations like Taurus. Then, the lower BF measured in clusters
is thought to result from the rapid disruption of primordial binaries,
through destructive gravitational encounters that occur on a time scale of
less than 1 Myr (e.g.,  Kroupa, Petr, \& MacCaughrean 1999). Since the rate
of gravitational encounters scales with the local stellar density, this
scenario conceivably accounts for the observed trend of lower binary
fractions in denser star forming regions (Patience \& Duch\^ene 2001), 
and it is further supported by the paucity of wide binaries observed in 
the ONC (Scally et al. 1999) and, more generally, in young open clusters 
(Ghez 2001).

This mode of binary formation is not exempt from difficulties, however. One
issue is whether the Taurus binaries can be regarded as representative 
of a universal population of primordial binaries. The frequency of primordial
binaries that is expected from the decay of small-N aggregates is of order of
BF$_p$ $\simeq$ 1 / (N-1), i.e., at most 50\% for N=3. This is
significantly lower than the BF measured in the Taurus association, which
amounts to $\geq$80\% for stars in the mass range $\sim 0.3-1.2 \msun$, with
little dependence on the primary mass (Leinert et al. 1993). This discrepancy 
could be solved if single fragments that were dynamically ejected from
protostellar aggregates had escaped from their birth place.  With typical
ejection velocities of 3-4 \kms (Sterzik \& Durisen 1995), they would be
located a few parsecs away from their birth site at an age of 2~Myr, i.e.,
a few degrees away from the Taurus stellar groups (Gomez et al.  1995).
A widely distributed population of X-ray emitting T Tauri stars has been 
detected with ROSAT over the Taurus cloud (Wichmann et al.  1996,
2000, Frink et al.  1997), but these stars do not seem to be preferentially
single (K\"ohler \& Leinert 1998), as would be expected if they were
escapers.

An intriguing possibility is that the ejected fragments are very low mass,
indeed substellar, objects (Sterzik \& Durisen 1999, Clarke \& Reipurth
2001) that might so far have escaped detection in the Taurus cloud.
Although the search for brown dwarfs in Taurus has been somewhat
disappointing (Luhman 2000), it has only concentrated on very limited areas
centered on the small Taurus stellar groups. A widely distributed
population of (single) substellar objects over the Taurus cloud could
reconcile the high BF frequency measured for Taurus {\it stars\/} with the
lower BF expected from the decay of small-N aggregates, which includes both
stellar and substellar fragments. In support of this hypothesis, we note
that current determinations of the substellar IMF do indicate that isolated
brown dwarfs are numerous in clusters (e.g.  Luhman et al.  2000, Moraux et
al. 2001) and appear to be preferentially single objects (Mart\'{\i}n et
al. 1999). If originally ejected from small-N aggregates, substellar
fragments may be more easily retained in the deep potential well of dense
clusters than in loose associations (de la Fuente Marcos \& de la Fuente
Marcos 2000), which might explain why they have not been found in the
central regions of Taurus.

Another aspect of the models that is challenged by the observations is
whether the lower BF of clusters compared to associations can be understood
as the mere result of the disruption of primordial binaries. Models that
describe the dynamical evolution of primordial binaries in young clusters,
starting from an initial distribution similar to the one observed in Taurus,
show that the destruction rate of wide primordial binaries
($\logp\sim$5-7) is a sensitive function of the initial stellar density
(e.g.,  Kroupa 1995; Kroupa, Aarseth \& Hurley 2000). Yet, all clusters
studied so far appear to harbour the same BF to within a few percent in this
\logp range.  The lack of dispersion in the BF measured for clusters is
then surprising, given that it is unlikely all clusters surveyed so far
have formed with precisely similar densities. For instance, the stellar
density in the Trapezium cluster is of order of 5 10$^4$ pc$^{-3}$
(McCaughrean \& Stauffer 1994) whereas, at a similar age, it is about 5
10$^3$ pc$^{-3}$ in IC 348 (Herbig 1998). If gravitational encounters
leading to the disruption of primordial binaries are the dominant mechanism
that yields a lower BF in clusters, one would expect to observe somewhat
different binary fractions between clusters themselves.

Hence, while scenarios of binary formation and evolution that assume an
initially large fraction of primordial binaries in all star forming regions,
followed by a rapid erosion of the binary population in dense clusters, have
recently become quite popular, it remains to be seen whether the
difficulties outlined above can be solved.

\subsection{Do local conditions impact on the binary formation process?}

As an alternative to a universal formation mechanism, it is probably too
early yet to rule out binary and, indeed, single star formation as the
direct outcome of cloud collapse and fragmentation, without going through
the transient episode of small-N protostellar aggregates. Unfortunately,
the theory and simulations of fragmentation are not yet predictive enough,
and the final product of protostellar collapse can depend sensitively on
initial conditions (see Bodenheimer et al.  2000 for a review), e.g., the
radial density profile of the parental cloud (Burkert et al.  1997), its
temperature (Sterzik and Durisen 1995), turbulence (Klein 2001), the
magnetic field (Boss 2001), etc.  The large scale environment, such as
cloud-cloud collisions, or other external impulsive processes, such as
supernova blasts, may also impact on the fragmentation process (Whitworth
2001). Hence, one might expect that different initial conditions in star
forming regions lead to significant variations in the properties of the
young stellar populations they harbour. 

Since the initial conditions that led to star formation in a given
molecular cloud are usually poorly known, it is somewhat difficult to
constrain this alternative mode of binary and single stars formation with
current observations. As noted above, however, one of the striking results
of the recent binary surveys is the quasi-universality of the BF in
clusters, which all appear to harbour the same fraction of solar-type, wide
(sep $\geq$ 20 AU) binaries to within the statistical uncertainties. This
suggests that the fragmentation process, if directly responsible for the
formation of binary systems, might not be as sensitive to local conditions
as numerical simulations tend to indicate. On the other hand, while the
results for cluster binaries are homogeneous and similar to those obtained
for the field binary population, the much larger BF observed in the Taurus
cloud would seem to suggest that gross variations in the local conditions
do impact on the fragmentation process.

In this respect, it is interesting to note that binary frequency is not the
only difference that exists between the stellar populations of the Orion
cluster and Taurus association. Significant differences have also been
found in the distribution of stellar angular momentum among their
low-mass T Tauri stars (Clarke \& Bouvier 2000), and in the distribution 
of their stellar masses, with Taurus harbouring apparently both
fewer high-mass stars and fewer very-low mass objects than Orion
(Hartmann\& Kenyon 1995, Luhman 2000). It is tempting to think that the
differences observed in the fundamental properties (mass and angular
momentum distributions, binary frequency) of the Orion and Taurus
populations are causally related and point to a common origin that reflects
intrinsically different modes of star formation in clusters and in
associations (e.g., Myers 1998, Williams et al. 2000, Motte \& Andr\'e
2001).

\section{Conclusion}

From an adaptive optics imaging survey of 149 G and K-type primaries of the
Praesepe cluster, we find that solar-type cluster members harbour the same
proportion of close visual binaries as do G-type field dwarfs. Long lived
open clusters, such as Praesepe, probably started their evolution as
extremely dense protostellar clusters. Yet, only about 10\% of the field
population is thought to result from the dissipation of such rich clusters.
At the other extreme, Taurus-like regions of distributed star formation
have very low star forming efficiencies. Hence, as recently advocated by
Adams \& Myers (2001), most field stars must have been born in stellar
groups which dissipate in a few million years, corresponding to initial
conditions somewhat intermediate between dense protostellar clusters
destined to become young open clusters and loose associations. The very
similar binary fraction measured for solar-type stars in young open
clusters (Praesepe, Pleiades, Alpha Per) and in the field thus suggest that
the formation and evolution of low-mass binaries is not very sensitive to
local conditions.

The main limitation of studies like the present one which aim at
constraining the star formation process through the investigation of young
binaries is that they have been mostly concerned with low-mass systems so
far and somewhat neglected higher mass binaries (see, however, Preibisch et
al.  1999, Garc\`{\i}a \& Mermilliod 2001, Duch\^ene et al. 2001). If
multiple fragmentation of collapsing clouds is the dominant mode for the
formation of multiple stellar systems, the mass distribution of fragments
in small protostellar groups may largely determine the resulting binary
frequency for a given primary mass. For instance, we argued above that the
high BF observed for T Tauri stars in Taurus might merely be the result of
neglecting a putative population of single brown dwarfs distributed over
the cloud. In regions where high mass stars are formed, such as in Orion,
more single ejected fragments would be of solar mass or so, thus resulting
in a lower binary fraction among low mass stars. The investigation of such
causal relationships between the fundamental properties of young stars,
e.g., between binary fraction and the mass function, requires the
consideration of the whole stellar population of the star forming region
with a complete census of multiple systems at all primary masses, which is
not available today for any star forming region.

A promising new way to better understand the formation of multiple systems
is to investigate, in different environments, extremely young stellar
objects still embedded in their natal cloud at the end of the protostellar
collapse. The high degree of multiplicity of such Class 0 and Class I
``protostellar'' sources starts to be revealed from high angular resolution
studies in the millimeter range (Looney, Mundy \& Welch 2000). The advent
of adaptive optics systems equipped with near-IR wavefront sensors on large
telescopes will now open the way to large scale surveys of embedded
protobinaries with a tenfold increase in angular resolution compared to
current millimeter studies, reaching separations as small as a few
astronomical units. Such studies will provide unprecedented details on the
fragmentation process at a very early stage of evolution of young systems,
before any significant dynamical evolution of protostellar systems has
occurred.

\begin{acknowledgements}
  We thank Jean-Luc Beuzit and Olivier Lai for reobserving in Nov.-Dec.
  1999 some suspected binaries with the same instrumentation and Isabelle
  Baraffe for computing and providing a 700 Myr isochrone from her models
  of low mass stars. We acknowledge useful discussions with Cathie Clarke
  and Pavel Kroupa on cluster dynamics and with Fr\'ed\'erique Motte on
  prestellar cores.
\end{acknowledgements}


\begin{thebibliography}{}
\bibitem[2000]{abt00}
Abt H.A., Wilmarth D.W. 2000, ApJ 521, 682
\bibitem[2001]{adams2001}
Adams F.C., Myers P.C. 2001, ApJ in press, astro-ph/0102039
\bibitem[1998]{baraffe98}
Baraffe I., Chabrier G., Allard F., Hauschildt P.H. 1998, AA 337, 403
\bibitem[1998]{bat98}
Bate M.R. 1998, ApJ 508, L95
\bibitem[2001]{bod01}
Bodenheimer P. 2001, in: The Formation of Binary Stars, ASP Conf. Ser.,
Vol.200, eds Zinnecker \& Mathieu, in press
\bibitem[2000]{bod00}
Bodenheimer P., Burkert A., Klein R.I., Boss A.P. 2000, in: Protostars \&
Planets IV, eds. Mannings et al., p.675
\bibitem[2001]{bos01}
Boss A.P. 2001, in: The Formation of Binary Stars, ASP Conf. Ser.,
Vol.200, eds Zinnecker \& Mathieu, in press
\bibitem[1997]{bouvier97}
Bouvier J., Rigaut F., Nadeau D. 1997, AA 323, 139
\bibitem[1997]{bur97}
Burkert A., Bate M.R., Bodenheimer P. 1997, MNRAS 289, 497
\bibitem[2001]{cla01}
Clarke C.J. 2001, in: The Formation of Binary Stars, ASP Conf. Ser.,
Vol.200, eds Zinnecker \& Mathieu, in press
\bibitem[2000]{cla00}
Clarke C.J., Bouvier J. 2000, MNRAS 319, 457
\bibitem[1991]{cla91}
Clarke C.J., Pringle J.E. 1991, MNRAS 249, 584
\bibitem[2001]{cla01}
Clarke C.J., Reipurth B. 2001, in press
\bibitem[1997]{fue97}
de la Fuente Marcos R. 1997, AA 322, 764
\bibitem[2000]{fue00}
de la Fuente Marcos R., de la Fuente Marcos C. 2000, ApSS 271, 127
\bibitem[1998]{doyon98}
Doyon R., Nadeau D., Vallee P., et al. 1998, in: Fowler A.M. (ed.), SPIE
Proc. 3354, Infrared Astronomical Instrumentation 
\bibitem[1999]{duc99}
Duch\^ene G. 1999, AA 341, 547
\bibitem[1999]{duc99}
Duch\^ene G., Bouvier J., Simon T. 1999, AA 343, 831
\bibitem[2001]{duc01}
Duch\^ene G., Eisl\"offel J., Simon T., Bouvier J., 2001, AA, submitted
\bibitem[1991]{dm91} 
Duquennoy A., Mayor M. 1991, AA 248, 485 (DM91)
\bibitem[1994]{dur94}
Durisen R.H., Sterzik M.F. 1994, AA 286, 84
\bibitem[1986]{dur86}
Durisen R.H., Gingold R.A., Tohline J.E., Boss A.p. 1986, ApJ 305, 281
\bibitem[1999]{eis01}
Eisl\"offel J., Simon T., Close L., Bouvier J. 2001, 11th Cambridge
workshop on Cool Stars, Stellar Systems and the Sun, ASP Conf. Ser.,
eds Garcia Lopez et al., Vol. 223, in press
\bibitem[1992]{fm92}
Fischer D.A., Marcy G.W. 1992, ApJ 396, 178
\bibitem[1997]{fri97}
Frink S., R\"oser S., Neuha\"user R., Sterzik M.F. 1997, AA 325, 613
\bibitem[2001]{gm01}
Garc\`{\i}a B., Mermilliod J.-C. 2001, AA 368, 122
\bibitem[2001]{ghe01}
Ghez A.M. 2001, in: The Formation of Binary Stars, ASP Conf. Ser.,
Vol.200, eds Zinnecker \& Mathieu, in press
\bibitem[1993]{ghe93}
Ghez A., Neugebauer G., Matthews K. 1993, AJ 106, 2005
\bibitem[1993]{gom93}
Gomez M., Hartmann L., Kenyon S.J., Hewett R. 1993, AJ 105, 1927
\bibitem[1998]{her98}
Herbig G.H. 1998, ApJ 497, 736
\bibitem[1991]{js91}
Jones B.F., Stauffer J.R. 1991, AJ 102, 1080
\bibitem[1995]{har95}
Kenyon S.J., Hartmann L.W. 1995, ApJS 101, 117
\bibitem[1927]{kw27}
Klein-Wassink W.J. 1927, Publ. Kapteyn Astr. Lab. no 41
\bibitem[2001]{kle01}
Klein R.I., Fisher R., McKee C.F. 2001, in: The Formation of Binary Stars,
ASP Conf. Ser., Vol.200, eds Zinnecker \& Mathieu, in press
\bibitem[2000]{kle00}
Klessen R.S., Burkert A. 2000, ApJS 128, 287
\bibitem[1998]{koh98}
K\"ohler R., Leinert Ch. 1998, AA 331, 977
\bibitem[1995]{kro95}
Kroupa P. 1995, MNRAS 277, 1507
\bibitem[2000]{kro00}
Kroupa P. 2000, in: Massive Stellar Clusters, ASP Conf. Ser. 211, eds
Lan\c{c}on \& Boily, p.233
\bibitem[2001]{kro01}
Kroupa P., Aaserth P., Hurley S.J. 2001, MNRAS 321, 699
\bibitem[1999]{kro99}
Kroupa P., Petr M.G, McCaughrean M.J. 1999, New Astron. 4, 495
\bibitem[1993]{lei93}
Leinert Ch., Zinnecker H., Weitzel N. et al. 1993, AA 278, 129
\bibitem[2000]{loo00}
Looney L.W., Mundy L.G., Welsch W.J. 2000, ApJ 529, 477
\bibitem[2000]{luh00}
Luhman K.L. 2000, ApJ 544, 1044
\bibitem[2000]{luh00}
Luhman K.L., Rieke G.H., Young E.T., et al. 2000, ApJ 540, 1016 
\bibitem[1994]{mmj94}
McCaughrean M.J., Stauffer J.R. 1994, AJ 108, 1382
\bibitem[1993]{mcd93}
McDonald J.M., Clarke C.J. 1993, MNRAS 262, 800
\bibitem[1995]{mcd95}
McDonald J.M., Clarke C.J. 1995, MNRAS 275, 671
\bibitem[2000]{mar00}
Mart\'{\i}n E.L., Brandner W., Bouvier J., et al. 2000, ApJ 543, 299
\bibitem[2000]{mat00}
Mathieu R.D, Ghez A.M., Jensen E.L.N., Simon M. 2000, in: Protostars \&
Planets IV, eds. Mannings et al., p.703
\bibitem[1999]{mermilliodwebda}
Mermilliod J.-C. 1999, Webda: http://obswww.unige.ch/ webda/webda.html
\bibitem[1999]{mermilliod99}
Mermilliod J.-C., Mayor M. 1999, AA 352, 479
\bibitem[1994]{mermilliod94}
Mermilliod J.-C., Duquennoy A., Mayor M. 1994, AA 283, 515
\bibitem[1990]{mermilliod90}
Mermilliod J.-C., Weis E.W., Duquennoy A., Mayor M. 1990, AA 235, 214
\bibitem[2001]{mor01}
Moraux E., Bouvier J., Stauffer J.R. 2001, AA 367, 211
\bibitem[2001]{mot01}
Motte F., Andr\'e P. 2001, AA 365, 440
\bibitem[1998]{mye98}
Myers P.C. 1998, ApJ 496, L109
\bibitem[1994]{nadeau94}
Nadeau D., Murphy D.C., Doyon R., Rowlands N. 1994, PASP 106, 909
\bibitem[2001]{pat01}
Patience J., Duch\^ene G. 2001, in: The Formation of Binary Stars, ASP
Conf. Ser., Vol.200, eds Zinnecker \& Mathieu, in press
\bibitem[1998]{pat98}
Patience J., Ghez A., Reid I., Weinberger A., Matthews K. 1998, AJ 115,
1972
\bibitem[1998]{pet98}
Petr M., Coud\'e du Foresto V., Beckwith S., Richichi A., McCaughrean
M. 1998, ApJ 500, 825
\bibitem[1999]{pr99}
Preibisch T., Balega Y., Hofmann K.-H., Weigelt G., Zinnecker H. 1999, NewA
4, 531
\bibitem[1994]{pro94}
Prosser C.F., Stauffer J.R., Hartmann L.W., et al. 1994, ApJ 421, 517
\bibitem[1998]{rigaut98} 
Rigaut F., Salmon D., Arsenault R., et al. 1998, PASP 110, 152
\bibitem[1999]{robichon99}
Robichon N., Arenou F., Mermilliod J.-C., Turon C. 1999, AA 345, 471
\bibitem[1996]{robin}
Robin A., Haywood M., Cr\'ez\'e M., Ojha D.K., Bienaym\'e O. 1996, AA 305,
125
\bibitem[1999]{sca99}
Scally A., Clarke C., McCaughrean M.J., 1999, MNRAS 306, 253
\bibitem[1999]{sim99}
Simon M., Close L.M., Beck T.L. 1999, AJ 117, 1375
\bibitem[1995]{sim95}
Simon M., Ghez A., Leinert Ch. et al. 1995, ApJ 443, 625
\bibitem[1995]{dur95}
Sterzik M.F., Durisen R.H. 1995, AA 304, L9
\bibitem[1998]{dur98}
Sterzik M.F., Durisen R.H.1998, AA 339, 95
\bibitem[1999]{dur99}
Sterzik M.F., Durisen R.H.  1999, in: Star Formation 1999,
ed. T.~Nakamoto, p.387 
\bibitem[1997]{tok97}
Tokovinin A. 1992, AA 256, 121
\bibitem[1933]{vl33}
Vanderlinden H.L. 1933, Etude de l'amas de Praesepe, Duculot Editeur
(Gembloux)
\bibitem[1993]{vandessel93}
Van Dessel A., Sinachopoulos D. 1993, AAS 100, 517
\bibitem[2001]{wit01}
Whitworth A.P. 2001, in: The Formation of Binary Stars, ASP Conf. Ser.,
Vol.200, eds Zinnecker \& Mathieu, in press
\bibitem[1996]{wic96}
Wichmann R., Krautter J., Schmitt J.H.M.M., et al. 1996, AA 312, 439
\bibitem[2000]{wic00}
Wichmann R., Torres G., Melo C.H.F., et al. 2000, AA 359, 181
\bibitem[2000]{wil00}
Williams J.P., Blitz L., McKee C.F. 2000, in: Protostars \& Planets IV,
eds. Mannings et al., p.97 
\end{thebibliography}
\end{document}